\documentclass[pdflatex,sn-basic]{sn-jnl}
\usepackage[latin1]{inputenc}
\usepackage[T1]{fontenc}
\usepackage{graphicx}
\usepackage{enumitem}
\usepackage{latexsym}
\usepackage{amssymb}
\usepackage{epsfig}
\usepackage{amsmath}
\usepackage{natbib}
\usepackage{float}
\allowdisplaybreaks 

\numberwithin{equation}{section}

\newcommand{\Var}{\mbox{Var}}
\def\eps{\varepsilon}
\def \R{\mathbb{R}}

\def \Cov{\mbox{Cov}}

\def\er{\mathbb{R}}

\def\e{\varepsilon}

\def\beq{\begin{eqnarray*}}
\def\eeq{\end{eqnarray*}}

\begin{document}

\title{\bf Goodness-of-fit testing for the error distribution in  functional linear models
}

\author*[1]{\textsc{ Natalie Neumeyer}}\email{natalie.neumeyer@uni-hamburg.de}
\author*[1]{\textsc{Leonie Selk}}\email{leonie.selk@uni-hamburg.de}

\affil*[1]{University of Hamburg, Department of Mathematics}

\abstract{We consider the error distribution in  functional linear models with scalar response and functional covariate. Different asymptotic expansions of the empirical distribution function and the empirical characteristic function based on estimated residuals under different model assumptions are discussed. The results are applied for simple and composite goodness-of-fit testing for the error distribution, in particular testing for normal distribution.}


\maketitle

\newtheorem{theo}{Theorem}[section]
\newtheorem{lemma}[theo]{Lemma}
\newtheorem{cor}[theo]{Corollary}
\newtheorem{rem}[theo]{Remark}
\newtheorem{prop}[theo]{Proposition}
\newtheorem{defin}[theo]{Definition}
\newtheorem{example}[theo]{Example}
\newtheorem{Assumption}{Assumption}


\subsection*{Acknowledgement.}  Charles University of Prague and University of Hamburg are partner universities. Since 1980 the DAAD (Deutscher Akademischer Austauschdienst) supports several projects, including some between the Department of Probability and Mathematical Statistics, Charles University of Prague, and the Department of Mathematics, University of Hamburg.
 This collaboration led to several joint articles by Marie Hu\v{s}kov\'{a} with Konrad Behnen and Georg Neuhaus. In 2011, Marie started working with us, and we are very thankful for the great collaboration we still have today. Marie's impressive expertise in multiple areas of mathematical statistics and her encouragement to collaborate have been very inspiring for us. We are also very grateful for her kind hospitality during our visits in Prague.

\section{Introduction}

In this paper we consider a functional linear model with scalar response $Y$ and functional covariate $X$, and we are especially interested in estimating the error distribution. Our aim is to test for parametric classes of error distributions, in particular for Gaussian errors.
Gaussian distribution is a typical assumption in functional linear regression models, see e.\,g.\ \cite{CaiEtal2018} or \cite{JamesEtal2009}. Often, other assumptions can be relaxed, when the error distribution is assumed to be normal, see e.\,g.\  \cite{ImaizumiKato2019} or \cite{HormannEtal2022}. 
To illustrate this we give some details of the mentioned papers: In \cite{CaiEtal2018} a functional linear model with scalar response and Gaussian errors is considered. Their aim is to estimate the slope function $\beta$, adaptive to unknown smoothness parameters. 
Similarly, \cite{JamesEtal2009} assume Gaussian errors and aim to estimate $\beta$ in a way that identifies regions where $\beta$ is zero and gives a simple structure over the remaining regions. They show that a faster rate of convergence for their $\hat\beta$ is obtained when, in addition to the Gaussianity of the independent errors, some extended errors that depend on $X$ follow a normal distribution as well.
\cite{ImaizumiKato2019} introduce confidence bands for the slope function $\beta$ under the assumption of Gaussian errors. To allow for other error distributions, they require, among others, higher moments of the parameters in the orthonormal basis decomposition of the functional covariate $X$.  
\cite{HormannEtal2022} estimate the distribution of a functional response $Y$ given a general predictor $X$ where the dependence between $Y$ and $X$ is given by a functional linear model.  Under the assumption of Gaussian errors, their method is simplified by modelling $Y$ conditioned on $X$ as a Gaussian process.

We consider tests based on the empirical distribution function of residuals as well as tests based on the empirical characteristic function of residuals. 
Expansions of the residual based empirical distribution function, and goodness-of-fit tests for the error distribution in  classical linear models with finite-dimensional covariates can be found in \cite{Koul2002} and \cite{NeumeyerEtal2006}.
Goodness-of-fit tests for the error distribution based on the empirical characteristic function were considered in \cite{HuskovaMeintanis2007}  for linear models with real valued covariates, in \cite{HuskovaMeintanis2009} for parametric regression models,  and in \cite{HuskovaMeintanis2010} for nonparametric regression models. 
There is more related literature with similar methods in other models, or other testing problems. E.\,g.\ \cite{HuskovaEtal2019}  compare the methods based on empirical characteristic function and on empirical distribution function to test for independence of innovations and past time series values for nonparametric autoregression models.

The paper is organized as follows. In Section \ref{sec-model} we introduce our model and present expansions  of the empirical distribution function  of residuals under different assumptions. Sections \ref{sec:cdf} and \ref{sec:ecf} contain the derivation of goodness-of-fit tests based on the empirical cdf and the empirical characteristic function, respectively, for simple as well as composite hypothesis on the the error distribution, in particular tests for Gaussian errors. Finite sample properties are shown in Section \ref{sec:simus}, including some simulations of a misspecified model. Section \ref{concluding remarks} concludes the paper with some discussion of related problems. 
Proofs of some auxiliary results are given in the appendix.

\section{Model, assumptions and expansion of the empirical cdf of residuals}\label{sec-model}

Let $\mathcal H$ be a separable Hilbert space with inner product $\langle \cdot,\cdot\rangle$, corresponding norm $\|\cdot\|$ and Borel-sigma field.  Let $(X_i,Y_i)$, $i=1,\dots,n$, be an independent sample of ($\mathcal{H}\times \mathbb{R}$)-valued random variables defined on the same probability space with probability measure $\mathbb{P}$. The data  are assumed to fulfill a functional linear model
\begin{equation}\label{model}
Y_i=\alpha+ \langle X_i,\beta\rangle+ \eps_i,\quad i=1,\dots,n,
\end{equation}
 with scalar response $Y_i$ and $\mathcal H$-valued covariate $X_i$, and with  parameters $\alpha\in\R$, $\beta\in \mathcal H$. The covariates $X_1,\dots,X_n$ are assumed to be independent and identically distributed, and the errors $\eps_1,\dots,\eps_n$ are independent, centred, identically distributed with cdf $F$, and independent of the covariates. 
 In this section we consider asymptotic expansions in different cases for the empirical distribution function of residuals. Those results will then be applied for goodness-of-fit testing of the error distribution in Section 3. They can also be applied for other hypotheses concerning the error distribution, e.g.\ symmetry, as we will discuss in Section 6. 
  
 Let $\hat\alpha$ and $\hat\beta$ be estimators of $\alpha$ and $\beta$, and define residuals $\hat\eps_i=Y_i-\hat\alpha-\langle X_i,\hat\beta\rangle$. The residual-based and error-based empirical cdfs are defined as 
\begin{equation*}
\hat F_{n}(z)=\frac{1}{n}\sum_{i=1}^{n} I\{\hat\eps_i\leq z\},\quad F_{n}(z)=\frac{1}{n}\sum_{i=1}^{n} I\{\eps_i\leq z\}.
\end{equation*}

We formulate some assumptions we will need for the asymptotic results.  We use the notations $\overline{X}_n=\frac1n \sum_{i=1}^n X_i$, $\overline{Y}_n=\frac1n \sum_{i=1}^n Y_i$, $\overline{\eps}_n=\frac1n \sum_{i=1}^n \eps_i$. Let $(X,Y)$ have the same distribution as $(X_1,Y_1)$, but be independent of the sample $(X_i,Y_i)$, $i=1,\dots,n$. 

\begin{itemize}
\item[\bf (M)] Let $(X_i,Y_i)$, $i=1,\dots,n$, be independent identically distributed data, fulfilling the functional linear model (\ref{model}), where $\eps_1,\dots,\eps_n$ are independent of $X_1,\dots,X_n$.
\item[\bf (e)] Let $\eps_1,\dots,\eps_n$ be independent and identically distributed with cdf $F$ that  is twice differentiable with  bounded density $F^\prime=f$ and bounded $f^\prime$.   
\item[\bf (X)] $E\|X\|^2<\infty$
\item[\bf (b)] $\mathbb{P}\big(\hat \beta-\beta\in\mathcal{B}\big)\to 1$ as $n\to\infty$ for 
a  class $\mathcal{B}\subset\mathcal H$ such that the function class
$\mathcal{F}= \left\{(x,e)\mapsto I\{e\leq v+\langle x,b\rangle\}\mid v\in\R, b\in \mathcal{B}\right\}$
is $P$-Donsker, where $P$ denotes the distribution of $(X_1,\eps_1)$. 
\end{itemize}

Based on the slope estimator $\hat\beta$ we recommend to apply the  estimator $\hat\alpha=\overline{Y}_n-\langle \overline{X}_n,\hat\beta\rangle$ for the intercept $\alpha=E[Y]-\langle E[X],\beta\rangle$. Then for the first asymptotic result we only need consistency of $\hat\beta$ in the sense $\|\hat\beta-\beta\|=o_\mathbb{P}(1)$. Note that by Cauchy Schwarz inequality and the law of large numbers one obtains also consistency of $\hat\alpha$,  
$$|\hat\alpha-\alpha|=|\overline{\varepsilon}_n-\langle \overline{X}_n,\hat\beta-\beta\rangle|\leq |\overline{\varepsilon}_n|+\|\overline{X}_n\|\cdot\|\hat\beta-\beta\|=o_\mathbb{P}(1).$$
In the literature on functional linear models often the intercept is ignored, i.e.\ $\alpha=0$ in our model (\ref{model}). For this reason we also consider the case $\alpha=\hat\alpha=0$. Interestingly one obtains a different asymptotic expansion than in the model with intercept, which is simpler in the case $E[X]=0$, but rather problematic for the case $E[X]\neq 0$. Thus we consider the following three cases (ab.1), (ab.2) and (ab.3), and discuss the different asymptotic results later.

\begin{itemize}
\item[\bf (ab.1)] $\hat\alpha=\overline{Y}_n-\langle \overline{X}_n,\hat\beta\rangle$,  $\|\hat\beta-\beta\|=o_\mathbb{P}(1)$.
\item[\bf (ab.2)] $\alpha=\hat\alpha=0$, $\|\hat\beta-\beta\|=o_{\mathbb{P}}(n^{-1/4})$, and $E[X]=0$.
\item[\bf (ab.3)] $\alpha=\hat\alpha=0$, $\|\hat\beta-\beta\|=o_{\mathbb{P}}(n^{-1/4})$, and $E[X]\neq 0$.
\end{itemize}

We formulate the  theorem which follows from results in \cite{NeumeyerSelk2025}.

\begin{theo}\label{theo1}
Under  assumptions (M), (e), (X), and (b), the following expansions of the residual empirical distribution function hold. 
\begin{itemize}
\item[(i)] If (ab.1) holds and $E[\eps_1^2]<\infty$, then 
$\hat F_n(z)=F_n(z)+f(z)\overline{\eps}_n+o_{\mathbb{P}}(n^{-1/2})$ uniformly in $z\in\mathbb{R}$.
\item[(ii)] If (ab.2) holds, then 
$\hat F_n(z)=F_n(z)+o_{\mathbb{P}}(n^{-1/2})$ uniformly in $z\in\mathbb{R}$.
\item[(iii)] If (ab.3) holds, then 
$\hat F_n(z)=F_n(z)+f(z)\langle E[X],\hat\beta-\beta\rangle +o_{\mathbb{P}}(n^{-1/2})$ uniformly in $z\in\mathbb{R}$.
\end{itemize} 
\end{theo}

\noindent {\bf Proof of Theorem \ref{theo1}.} The assertions follow from Theorem 2.1 and Section 6 in \cite{NeumeyerSelk2025}. We give a few details that are on the one hand needed for explanations in Remark \ref{rem1}, and on the other hand to explain the expansions under different  assumptions. 
In the proof of Theorem 2.1 in the mentioned paper it was shown under assumptions (b), $E\|X\|<\infty$, $|\hat\alpha-\alpha|+\|\hat\beta-\beta\|=o_{\mathbb{P}}(1)$, and less restrictive assumptions than (e), that 
$$\hat F_n(z)=F_n(z)+R_n(z)+o_{\mathbb{P}}(n^{-1/2})$$ 
uniformly in $z\in\R$, where
\begin{eqnarray}\label{Rn}
R_n(z) &=& E_X[F(z+\hat\alpha-\alpha+\langle X,\hat\beta-\beta\rangle)]-F(z).
\end{eqnarray} 
Here $E_X$ denotes the expectation with respect to $X$, which has the same distribution as $X_i$, but is independent of $\hat\alpha$, $\hat\beta$.  
This derivation  arises from the presentation
$$\sqrt{n}(\hat F_n(z)-F_n(z)-R_n(z))=H_n(z,\hat\alpha-\alpha,\hat\beta-\beta)$$
for the process
\begin{eqnarray*}
H_n(z,a,b)
&=&\frac{1}{n^{1/2}}
\sum_{i=1}^n (I\{\eps_i\leq z+a+\langle X_i,b\rangle\}-I\{\eps_i\leq z\}\\
&&\qquad\qquad{}-E[I\{\eps_i\leq z+a+\langle X_i,b\rangle\}-I\{\eps_i\leq z\}])
\end{eqnarray*}
($z,a\in\mathbb{R}$, $b\in\mathcal{B}$)
that is asymptotically stochastically equicontinuous and thus converges to zero if inserting $a=\hat\alpha-\alpha$ and $b=\hat\beta-\beta$.

\medskip

\noindent (i) If (ab.1) holds the remainder term (\ref{Rn}) is equal to 
\begin{eqnarray*}
R_n(z) &=& E_X[F(z+\overline{\eps}_n-\langle \overline{X}_n-X,\hat\beta-\beta\rangle) -F(z)]\\
&=& f(z)\left(\overline{\eps}_n-\langle \overline{X}_n-E[X],\hat\beta-\beta\rangle\right) +r_n(z)
\end{eqnarray*}
by a Taylor expansion. By Cauchy-Schwarz inequality, the assumption $\|\hat\beta-\beta\|=o_{\mathbb{P}}(1)$, and  $n^{1/2}\|\overline{X}_n-E[X]\|=O_{\mathbb{P}}(1)$ by the central limit theorem in Hilbert spaces, one obtains
\begin{eqnarray*}
R_n(z) &=&  f(z)\overline{\eps}_n+o_{\mathbb{P}}(n^{-1/2}) +r_n(z),
\end{eqnarray*}
and the remainder term from the Taylor expansion
$$r_n(z)=E_X\left[\frac{f^\prime(\xi_n)}{2}\left(\overline{\eps}_n-\langle \overline{X}_n-E[X],\hat\beta-\beta\rangle\right)^2\right]=o_{\mathbb{P}}(n^{-1/2})$$
with the same argument using boundedness of $f^\prime$ and the central limit theorem for $\overline{\eps}_n$ and $\overline{X}_n-E[X]$.

\medskip

\noindent (ii), (iii) Setting $\alpha=\hat\alpha=0$ in (\ref{Rn}) under conditions (ab.2) or (ab.3) one obtains the result immediately from a Taylor expansion of $F$, 
\begin{eqnarray*}
R_n(z)  
&=& f(z)\langle E[X],\hat\beta-\beta\rangle +
E_X\left[\frac{f^\prime(\xi_n)}{2}\langle X,\hat\beta-\beta\rangle^2\right]
\end{eqnarray*}
and by boundedness of $f^\prime$, Cauchy-Schwarz inequality and conditions (ab.2) or (ab.3) the second term is of order $o_{\mathbb{P}}(n^{-1/2})$. 
\hfill $\Box$

\bigskip


\begin{rem}\label{rem1}\rm
The asymptotics in case (ab.3) in Theorem \ref{theo1}(iii) are problematic because $\langle E[X],\beta\rangle$ typically has a slower rate of convergence than $n^{-1/2}$ as $F_n-F$ (see \cite{CardotEtal2007}, \cite{YeonEtal2023}, \cite{ShangCheng2015}), and thus it will dominate the asymptotic distribution. We recommend to use a model with intercept and an estimator as in the  condition (ab.1). Even if the asymptotic result under (ab.2) (Theorem \ref{theo1}(ii)) is simpler, the  assumptions are rather strict. 
Models without intercept,  and with $E[X]=0$ are often used in the literature, but one has to be aware that it only works for $E[Y]=0$.  Often it is argued that one can apply the results for estimators based on a model $Y=\langle X,\beta\rangle+\eps$ with $E[X]=0$ for centred observations and centred covariates. This means that the real model is
$$Y_i-E[Y]=\langle X_i-E[X],\beta\rangle+\eps_i$$
and one applies the procedures for
$$Y_i-\overline{Y}_n\approx\langle X_i-\overline{X}_n,\beta\rangle +\eps_i,$$
 but in our context this should not be done as it changes the asymptotic distribution. One might suggest to obtain the result as in case (ab.2), but in fact one obtains the result as in case (ab.1) as can be derived as follows. 
 The residuals here have the form
 $$\hat\eps_i=Y_i-\overline{Y}_n-\langle X_i-\overline{X}_n,\hat\beta\rangle
 =\eps_i-\overline{\eps}_n-\langle X_i-\overline{X}_n,\hat\beta-\beta\rangle$$
 and as in the proof of Theorem \ref{theo1} one has
 $$\sqrt{n}(\hat F_n(z)-F_n(z)-R_n(z))=H_n(z,\overline{\eps}_n,\overline{X}_n,\hat\beta-\beta)$$
for the process
\begin{eqnarray*}
H_n(z,a,c,b)
&=&\frac{1}{n^{1/2}}
\sum_{i=1}^n (I\{\eps_i\leq z+a+\langle X_i-c,b\rangle\}-I\{\eps_i\leq z\}\\
&&\qquad\qquad{}-E[I\{\eps_i\leq z+a+\langle X_i-c,b\rangle\}-I\{\eps_i\leq z\}])
\end{eqnarray*}
($z,a,c\in\mathbb{R}$, $b\in\mathcal{B}$), which converges to zero for $a=\overline{\eps}_n=o_{\mathbb{P}}(1)$ and $b=\hat\beta-\beta$ with $\|\hat\beta-\beta\|=o_{\mathbb{P}}(1)$ (for each $c$). The remainder term is then
$$R_n(z)=E_X[F(z+\overline{\eps}_n+\langle X-\overline{X}_n,\hat\beta-\beta\rangle)]-F(z),$$
the same as in the proof of Theorem \ref{theo1}(i) in case (ab.1).

\end{rem} 

\bigskip

For the first two cases we now formulate the result of weak convergence. Under (ab.1) we obtain 
$$\sqrt{n}(\hat F_n(z)-F(z))=\frac{1}{\sqrt{n}}\sum_{j=1}^n (I\{\eps_j\leq z\}-F(z)+f(z)\eps_j)+o_{\mathbb{P}}(1)$$
uniformly in $z$, and for the dominating term one obtains weak convergence by classical empirical process results. In case (ab.2) one just has to consider the empirical process of true errors as dominating term. 

\begin{cor}\label{cor-1}
Under assumptions (M), (e), (X), and (b) we obtain the following limit processes. 
\begin{itemize}
\item[(i)] If (ab.1) holds and $\sigma^2=\Var(\eps_1)\in (0,\infty)$, then the process 
$\sqrt{n}(\hat F_n-F)$ converges weakly in $\ell^\infty(\mathbb{R})$ to a centred Gaussian process $G$ with covariance \rm
\begin{eqnarray*}
\Cov(G(y),G(z))&=&F(y\wedge z)-F(y)F(z)+f(y)E[\eps I\{\eps\leq z\}]\\
&&{}+f(z)E[\eps I\{\eps\leq y\}]+\sigma^2 f(y)f(z).
\end{eqnarray*}\it
\item[(ii)] If (ab.2) holds,  
then the process 
$\sqrt{n}(\hat F_n-F)$ converges weakly in $\ell^\infty(\mathbb{R})$ to $B\circ F$, where $B$ denotes a standard Brownian bridge on $[0,1]$. 
\end{itemize} 
\end{cor}

\noindent In the next section those results are applied for goodness-of-fit testing.

\section{Goodness-of-fit testing based on empirical cdf}\label{sec:cdf}

\subsection{Simple hypothesis}\label{gof-edf-simple}

For goodness-of-fit testing we first consider the simple null hypothesis $H_0:F=F_0$ for some fixed cdf $F_0$. Assume that $F_0$ is twice differentiable with bounded density $F_0'=f_0$ and  bounded $f_0'$. The test statistic is a continuous function of the process $\sqrt{n}(\hat F_n-F_0)$, e.g.\ a Kolmogorov-Smirnov distance $\sqrt{n}\sup_{z\in\mathbb{R}}|\hat F_n(z)-F_0(z)|$
or Cram\'er-von Mises statistic
$n\int (\hat F_n-F_0)^2\,dF_0$,
and one can apply directly Corollary \ref{cor-1} and the continuous mapping theorem 
to obtain the limit distribution in settings (ab.1) and (ab.2).

\subsection{Composite hypothesis}\label{gof-normal}

 Our aim is to test for a finite-dimensional parametric class of the error distribution, i.e.\ the null hypothesis
 $$H_0: F\in\mathcal{F}=\{F_\vartheta\mid\vartheta\in\Theta\}.$$
 Test statistics can be based on the process $\sqrt{n}(\hat F_n-F_{\hat\vartheta})$, but due to the parameter estimation $\hat\vartheta$ the asymptotic distribution will be more complex than for the simple hypothesis. As a special case we consider testing for normal distribution with variance $\vartheta^2$, i.e.\ $F_\vartheta(\cdot)=\Phi(\frac{\cdot}{\vartheta})$, $\vartheta\in\R^+$, applying the residual-based Cram\'er-von Mises test
$$T_{n,D}=n\int (\hat F_n-F_{\hat\vartheta})^2\,dF_{\hat\vartheta}$$
with 
$\hat\vartheta^2=\frac{1}{n}\sum_{i=1}^n\hat\eps_i^2$. Note that we obtain the expansion $\hat\vartheta^2=\frac{1}{n}\sum_{i=1}^n \eps_i^2+r_n$,
where the remainder term can be upper bounded by 
\begin{eqnarray*}
|r_n|&\leq& 2|\overline{\eps}_n||\hat\alpha-\alpha|+2\left\|\frac{1}{n}\sum_{i=1}^n \eps_iX_i\right\|\|\hat\beta-\beta\|+(\hat\alpha-\alpha)^2+ \frac{1}{n}\sum_{i=1}^n \|X_i\|^2\|\hat\beta-\beta\|^2\\
&=& o_{\mathbb{P}}(n^{-1/2}),
\end{eqnarray*}
where we need assumption (X), condition (ab.1') formulated below, and law of large numbers and central limit theorem in Hilbert spaces. Further $E[\eps_1^2]<\infty$ is needed which holds under the null hypothesis of  a normal distribution. 

\begin{itemize}
\item[\bf (ab.1')] $\hat\alpha=\overline{Y}_n-\langle \overline{X}_n,\hat\beta\rangle$,  $\|\hat\beta-\beta\|=o_{\mathbb{P}}(n^{-1/4})$.
\end{itemize}
Note that under this condition one also obtains $|\hat\alpha-\alpha|=|\overline{\eps}_n-\langle \overline{X}_n,\hat\beta-\beta\rangle|=o_{\mathbb{P}}(n^{-1/4})$.

Under those assumptions we obtain 
\begin{eqnarray*}
\hat\vartheta^2&=&\frac{1}{n}\sum_{i=1}^n \hat\eps_i^2=\frac{1}{n}\sum_{i=1}^n \eps_i^2+ o_{\mathbb{P}}(n^{-1/2})=\vartheta^2+ O_{\mathbb{P}}(n^{-1/2})
\end{eqnarray*}
if fourth error moments exists, which holds under the null hypothesis. It then holds that
\begin{equation}\label{var-expansion}
\hat\vartheta-\vartheta=\frac{1}{2\vartheta}\frac{1}{n}\sum_{i=1}^n (\eps_i^2-\vartheta^2)+o_{\mathbb{P}}(n^{-1/2}).
\end{equation}
One can write the test statistic as
$$T_{n,D}=n\int (\hat F_n(\hat\vartheta y)-\Phi(y))^2 \, d\Phi(y)$$
and to derive the asymptotic distribution under $H_0$ we use the same method as in the proof of Theorem \ref{theo1}. Note that 
$\hat \eps_i\leq \hat\vartheta y$ is equivalent to
$$\eps_i\leq \hat\vartheta y+\hat\alpha-\alpha+\langle X_i,\hat\beta-\beta\rangle$$ 
and thus
 $$\sqrt{n}\left(\hat F_n(\hat\vartheta y)-F_n(\vartheta y)-R_n(y)\right)=H_n\Big(y,\vartheta,\frac{\hat\vartheta}{\vartheta},\hat\alpha-\alpha,\hat\beta-\beta\Big)$$
for the process
\begin{eqnarray*}
H_n(y,\vartheta,c,a,b)&=&\frac{1}{n^{1/2}}
\sum_{i=1}^n \Big(I\left\{\eps_i\leq c\vartheta y+a+\langle X_i,b\rangle\right\}-I\{\eps_i\leq \vartheta y\}\\
&&{}\qquad-E\left[I\left\{\eps_i\leq c\vartheta y+a+\langle X_i,b\rangle\right\}-I\{\eps_i\leq \vartheta y\}\right]\Big)
\end{eqnarray*}
($y,c,a\in\mathbb{R}$, $\vartheta\in\R^+$, $b\in\mathcal{B}$), which due to the Donsker property in assumption (b) converges to zero for $a=\hat\alpha-\alpha=o_{\mathbb{P}}(1)$, $c=\frac{\hat\vartheta}{\vartheta}=1+o_{\mathbb{P}}(1)$ and $b=\hat\beta-\beta$ with $\|\hat\beta-\beta\|=o_{\mathbb{P}}(1)$. The remainder term is then
\begin{eqnarray*}
R_n(z)&=&E_X\Big[\Phi\Big(\frac{\hat\vartheta}{\vartheta}y+\frac{\hat\alpha-\alpha}{\vartheta}+\frac{1}{\vartheta}\langle X,\hat\beta-\beta\rangle\Big)-\Phi(y)\Big]\\
&=& \phi(y)\Big(\frac{\hat\vartheta-\vartheta}{\vartheta}y+\frac{\hat\alpha-\alpha}{\vartheta}+\frac{1}{\vartheta}\langle E[X],\hat\beta-\beta\rangle\Big)+o_{\mathbb{P}}(n^{-1/2})\\
&=& \frac{\phi(y)}{\vartheta}\Big(\hat\alpha-\alpha+\langle E[X],\hat\beta-\beta\rangle\Big)+\frac{y\phi(y)}{2\vartheta^2}\frac1n \sum_{i=1}^n (\eps_i^2-\vartheta^2)
+o_{\mathbb{P}}(n^{-1/2}),
\end{eqnarray*}
where we use the notation $\Phi'=\phi$, and applied (\ref{var-expansion}).


In the simple case (ab.1') we can further derive
\begin{eqnarray*}
R_n(z)&=& \frac{\phi(y)}{\vartheta}\overline{\eps}_n+\frac{y\phi(y)}{2\vartheta^2}\frac1n \sum_{i=1}^n (\eps_i^2-\vartheta^2)
+o_{\mathbb{P}}(n^{-1/2}),
\end{eqnarray*}
whereas in the restrictive case (ab.2) it holds $\alpha=\hat\alpha=0$ as well as $E[X]=0$ and thus
\begin{eqnarray*}
R_n(z)&=& \frac{y\phi(y)}{2\vartheta^2}\frac1n \sum_{i=1}^n (\eps_i^2-\vartheta^2)
+o_{\mathbb{P}}(n^{-1/2}).
\end{eqnarray*}
From those expansions we 
 directly obtain the following result.

\begin{theo}\label{theo2}
Assume (M), (X), (b) under the null hypothesis of centred normally distributed iid errors,  with the notation $\tilde\eps_i=\eps_i/\vartheta\sim N(0,1)$. 
\begin{itemize}
\item[(i)] In case (ab.1') $$\hat F_n(\hat\vartheta y)-\Phi(y) =
\frac{1}{n}\sum_{i=1}^n \left(I\{\tilde\eps_i\leq y\}-\Phi(y)+\tilde\eps_i\phi(y)+(\tilde\eps_i^2-1)\frac{y\phi(y)}{2}\right)+o_{\mathbb{P}}(\frac{1}{\sqrt{n}})$$
holds uniformly in $y\in\mathbb{R}$, and $T_{n,D}$
 converges in distribution to 
$$T_{D_1}=\int_0^1 (G_{D_1}(t))^2\,dt$$
for a centred Gaussian process $G_{D_1}$ with covariance function
\rm
$$\Cov(G_{D_1}(s),G_{D_1}(t))=s\wedge t-st-\phi(\Phi^{-1}(s))\phi(\Phi^{-1}(t))\big(1+\frac 12\Phi^{-1}(s)\Phi^{-1}(t)\big).$$
\it
\item[(ii)] In case (ab.2)
$$\hat F_n(\hat\vartheta y)-\Phi(y) =
\frac{1}{n}\sum_{i=1}^n \left(I\{\tilde\eps_i\leq y\}-\Phi(y)+(\tilde\eps_i^2-1)\frac{y\phi(y)}{2}\right)+o_{\mathbb{P}}(\frac{1}{\sqrt{n}})$$
holds uniformly in $y\in\mathbb{R}$, and $T_{n,D}$
 converges in distribution to 
$T_{D_2}=\int_0^1  (G_{D_2}(t))^2\,dt$
for a centred Gaussian process $G_{D_2}$ with covariance function
\rm
$$\Cov(G_{D_2}(s),G_{D_2}(t))=s\wedge t-st-\frac 12\phi(\Phi^{-1}(s))\phi(\Phi^{-1}(t))\Phi^{-1}(s)\Phi^{-1}(t).$$
\it
\end{itemize}

\end{theo}

\medskip

In both cases the tests are  consistent and asymptotically distribution-free. However the limit distributions are different and according to Remark \ref{rem1} we recommend to use the model with intercept $\alpha$ and case (ab.1'), instead of (ab.2) without intercept. This will also be discussed in the simulation section in a misspecification setting. 
 Critical values can be calculated via Monte Carlo simulations of the processes $G_{D_1}$ and $G_{D_2}$.  In order to facilitate the reader's use of the tests, simulated asymptotic critical values $c_\alpha$ are shown in Tables \ref{tab1} and \ref{tab2}.
\begin{table}[h]
\small
\begin{tabular}{| l | l | l | l | l | l | }
\hline
nominal level  $\alpha$ &$0.15$&$0.1$&$0.05$&$0.025$&$0.01$\\
\hline
critical value $c_\alpha$ &$ 0.089$&$ 0.102 $&$ 0.125 $&$ 0.147 $&$ 0.177 $\\
\hline
\end{tabular}
\caption{\sl Asymptotic critical values for the Cram\'{e}r-von Mises test for normality of the errors in the functional linear model in case (ab.1').}\label{tab1} 
\small
\begin{tabular}{| l | l | l | l | l | l | }
\hline
nominal level  $\alpha$ &$0.15$&$0.1$&$0.05$&$0.025$&$0.01$\\
\hline
critical value $c_\alpha$ &$ 0.261$&$ 0.324 $&$ 0.437 $&$ 0.560 $&$ 0.729 $\\
\hline
\end{tabular}
\caption{ \label{tab2} \sl Asymptotic critical values for the Cram\'{e}r-von Mises test for normality of the errors in the functional linear model under condition (ab.2).}
\end{table}
\\
Now the tests can be applied by rejecting the null hypothesis of Gaussian errors when $T_{n,D}>c_\alpha$.\\
Note that in \cite{NeumeyerSelk2013}  the asymptotic covariance of a process with the same asymptotic expansion as in Theorem \ref{theo2}(i) has a slightly different form due to a calculation error. The formula given in the paper at hand is the correct one.

\section{Goodness-of-fit testing based on empirical characteristic function}\label{sec:ecf}

In addition to the tests based on the empirical cdf, we consider the empirical characteristic function (ecf)
$$\hat\varphi_n(t)=\frac 1n\sum_{j=1}^n\exp(it\hat\e_j)$$
with residuals $\hat\e_j=Y_j-\hat\alpha-\langle X_i,\hat\beta\rangle$, $j=1,\dots,n$,
as a basis for goodness-of-fit testing.

\subsection{Simple hypothesis}

First, we want to test the simple hypothesis like
$H_0: F=F_0$. Let $\eps_0$ be a random variable with cdf $F_0$ and characteristic function $\varphi_0(t)=E[e^{it\eps_0}]=\varphi_{0,1}(t)+i\varphi_{0,2}(t)$ with $\varphi_{0,1}(t)=E[\cos(\eps_0 t)]$, $\varphi_{0,2}(t)=E[\sin(\eps_0 t)]$.
A test statistic based on the empirical characteristic function is in this case
$$ T_{n,C_1}=n\int |\hat\varphi_n(t)-\varphi_0(t)|^2W(t)dt,$$
where $W$ is some non-negative symmetric weight function, satisfying $\int t^4W(t)dt<\infty$, e.\,g.\ $W(t)=\exp(-\frac 12t^2)$, which we used in the simulations (Section \ref{sec:simus}). 

Note that, due to symmetry properties of sine and cosine, we get
$$T_{n,C_1}=n\int\Big(\frac 1n\sum_{j=1}^n \sin(t\hat\e_j) +\cos(t\hat\e_j)-\varphi_{0,1}(t)-\varphi_{0,2}(t)\Big)^2W(t)dt.$$
In the case (ab.1') we need a higher moment condition for the covariates.
\begin{itemize}
\item[\bf (X2)] $E\|X\|^4<\infty$
\end{itemize}

\begin{theo}\label{theo3}
Under assumptions (M) and (e) under the null hypothesis $H_0:F=F_0$ we obtain the following limit distributions.
\begin{itemize}
\item[(i)] In case (ab.1') under assumption (X2) and $E[\eps_1^2]<\infty$,
\begin{eqnarray*}
T_{n,C_1}&
=& n\int \Big(\frac 1n\sum_{j=1}^n \sin(t\e_j)-E[\cos(t\e_1)]t\e_j+\cos(t\e_j)+E[\sin(t\e_1)]t\e_j\\
&&{}\qquad -\varphi_{0,1}(t)-\varphi_{0,2}(t)\Big)^2W(t)dt+ o_{\mathbb{P}}(1),
\end{eqnarray*}
 and $T_{n,C_1}$ converges in distribution to $T_{C_{1,1}}=\int (G_{C_{1,1}}(t))^2 W(t)dt$ for a centred Gaussian process $G_{C_{1,1}}$ with covariance function
\rm
$$\Cov(G_{C_{1,1}}(s),G_{C_{1,1}}(t))=\Cov(Z_1(s),Z_1(t))$$
\it
where $Z_1(s)=\sin(s\e_0)-\varphi_{0,1}(s)s\e_0+\cos(s\e_0)+\varphi_{0,2}(s)s\e_0$.
\item[(ii)] In case (ab.2) under assumption (X)
$$T_{n,C_1}=\ n\int \Big(\frac 1n\sum_{j=1}^n \sin(t\e_j)+\cos(t\e_j)-\varphi_{0,1}(t)-\varphi_{0,2}(t)\Big)^2W(t)dt+ o_{\mathbb{P}}(1),$$
which converges in distribution to $T_{C_{1,2}}=\int (G_{C_{1,2}}(t))^2 W(t)dt$ for a Gaussian process $G_{C_{1,2}}$ with covariance function
\rm
$$\Cov(G_{C_{1,2}}(s),G_{C_{1,2}}(t))=\Cov(Z_2(s),Z_2(t))$$
\it
where $Z_2(s)=\sin(s\e_0)+\cos(s\e_0)$.
\end{itemize}

\end{theo}

\noindent {\bf Proof of Theorem \ref{theo3}.} We present the main part of the proof here. Some auxiliary results can be found in the appendix. 
Note that by Taylor's expansion
\begin{equation}\label{expansion-1}
\frac 1n\sum_{j=1}^n \sin(t\hat\e_j) =\frac 1n\sum_{j=1}^n \sin(t\e_j)+\frac 1n\sum_{j=1}^n\cos(t\e_j)t(\hat\eps_j-\eps_j)+r_{n,1}(t),
\end{equation}
where
the remainder term fulfills
$$n\int (r_{n,1}(t))^2W(t)\,dt\leq \int t^4W(t)\,dt\left(\frac{1}{\sqrt{n}}\sum_{j=1}^n (\hat\eps_j-\eps_j)^2\right)^2=o_{\mathbb{P}}(1)$$
by Lemma \ref{lem-1}. 

\medskip

\noindent (i) In case (ab.1') it holds  that 
\begin{eqnarray}\nonumber
\hat\eps_j-\eps_j&=&\alpha-\hat\alpha+\langle X_j,\beta-\hat\beta\rangle=-\overline{\eps}_n-\langle X_j-\overline{X}_n,\hat\beta-\beta\rangle\\
&=& -\overline{\eps}_n-\langle X_j-E[X],\hat\beta-\beta\rangle +\langle \overline{X}_n-E[X],\hat\beta-\beta\rangle\label{expansion-hateps}
\end{eqnarray}
and  for the second sum in (\ref{expansion-1}) we obtain
\begin{eqnarray}\label{expansion-2}
\frac 1n\sum_{j=1}^n\cos(t\e_j)t(\hat\eps_j-\eps_j)
&=& -\overline{\eps}_n\frac 1n\sum_{j=1}^n\cos(t\e_j)t-r_{n,2}(t)+r_{n,3}(t)\\
&=&{}-\overline{\eps}_nE[\cos(t\e_1)]t-r_{n,2}(t)+r_{n,3}(t)- r_{n,4}(t)
\label{expansion-3}
\end{eqnarray}
with 
\begin{eqnarray*}
r_{n,2}(t) &=& \frac 1n\sum_{j=1}^n\cos(t\e_j)t\langle X_j-E[X],\hat\beta-\beta\rangle\\
r_{n,3}(t) &=& \frac 1n\sum_{j=1}^n\cos(t\e_j)t\langle \overline{X}_n-E[X],\hat\beta-\beta\rangle\\
r_{n,4}(t) &=& \overline{\eps}_n\frac 1n\sum_{j=1}^nt(\cos(t\e_j)-E[\cos(t\e_1)])
\end{eqnarray*}
and
$n\int (r_{n,\ell}(t))^2W(t)\,dt=o_{\mathbb{P}}(1)$ for $\ell=2,3,4$ by Lemmas \ref{lem-2} and \ref{lem-2b}, $\overline\e_n=O_{\mathbb{P}}(n^{-1/2})$, and Lemma \ref{lem-3}. 
Analogous expansions as in (\ref{expansion-1})--(\ref{expansion-3}) hold for $\frac 1n\sum_{j=1}^n\cos(t\hat\e_j)$ and from this we obtain assertion (i).


\medskip

\noindent (ii)
In case (ab.2)
we have
\begin{equation}\label{expansion-hateps-2}
\hat\eps_j-\eps_j=\langle X_j,\beta-\hat\beta\rangle=-\langle X_j-E[X],\hat\beta-\beta\rangle
\end{equation}
due to the assumption $E[X]=0$. 
As in (\ref{expansion-1}) and (\ref{expansion-2}) we can write
\begin{align*}
\frac 1n\sum_{j=1}^n \sin(t\hat\e_j) =&\ \frac 1n\sum_{j=1}^n \sin(t\e_j)+r_{n,1}(t)-r_{n,2}(t)
\end{align*}
and the assertion (ii) follows from Lemma \ref{lem-1}
and \ref{lem-2}.
\hfill $\Box$

\bigskip

When testing for standard normal errors, $\varphi_0(t)=\exp(-\frac 12 t^2)=\varphi_{0,1}(t)$, $\varphi_{0,2}(t)=0$, we obtain 
$$\Cov(G_{C_{1,1}}(s),G_{C_{1,1}}(t))=\exp\big(-\frac 12(s-t)^2\big)-(st+1)\exp\big(-\frac 12(s^2+t^2)\big)$$
and 
$$\Cov(G_{C_{1,2}}(s),G_{C_{1,2}}(t))=\exp\big(-\frac 12(s-t)^2\big)-\exp\big(-\frac 12(s^2+t^2)\big)$$
Critical values for $T_{C_{1,1}}$ and $T_{C_{1,2}}$ calculated via Monte Carlo simulations are tabled in Tables \ref{tab3} and \ref{tab4}, for the reader's ease of use of the tests. We reject the null hypothesis of a standard normal distribution if $T_{n,C_1}> c_\alpha$.
\medskip

\begin{table}[h]
\small
\begin{tabular}{| l | l | l | l | l | l | }
\hline
nominal level  $\alpha$ &$0.15$&$0.1$&$0.05$&$0.025$&$0.01$\\
\hline
critical value $c_\alpha$ &$ 1.052$&$ 1.272 $&$ 1.647 $&$ 2.036 $&$ 2.462 $\\
\hline
\end{tabular}
\caption{ \label{tab3} \sl Asymptotic critical values for the test for standard normal errors based on the empirical characteristic function with $W(t)=\exp(-\frac 12t^2)$ in the functional linear model under condition (ab.1').}
\small
\begin{tabular}{| l | l | l | l | l | l | }
\hline
nominal level  $\alpha$ &$0.15$&$0.1$&$0.05$&$0.025$&$0.01$\\
\hline
critical value $c_\alpha$ &$ 1.891$&$ 2.277 $&$ 2.932 $&$ 3.601 $&$ 4.603$\\
\hline
\end{tabular}
\caption{ \label{tab4} \sl Asymptotic critical values for the test for standard normal errors based on the empirical characteristic function with $W(t)=\exp(-\frac 12t^2)$ in the functional linear model under condition (ab.2).}
\end{table}

\subsection{Composite hypothesis}

For testing for a parametric class $H_0:F\in\{F_\vartheta\mid\vartheta\in\Theta\}$ one can use an $L^2$-distance between $\hat\varphi_n$ and the estimated characteristic function $\varphi_{\hat\vartheta}$, where $\hat\vartheta$ is a consistent estimator under $H_0$, and $\varphi_\vartheta$ is the notation for the characteristic function corresponding to the cdf $F_\vartheta$. We consider in detail again the test for normal distribution, i.e.\
$$H_0: F\in\Big\{\Phi\Big(\frac{\cdot}{\vartheta}\Big),\vartheta\in\er^+\Big\}.$$
Note that $\varphi_\vartheta(t)=\exp(-\frac12 \vartheta^2t^2)$. With the standard normal characteristic function $\varphi_1(t)=\varphi_\vartheta(t/\vartheta)=\exp(-\frac12 t^2)$ some suitable test statistic is
\begin{align*}
T_{n,C_2}=&\ \int |\hat\varphi_n(t/\hat\vartheta)-\varphi_1(t)|^2W(t)dt\\
=&\ n\int\Big(\frac 1n\sum_{j=1}^n \sin(t\hat\e_j/\hat\vartheta) +\cos(t\hat\e_j/\hat\vartheta)-\varphi_1(t)\Big)^2W(t)dt,
\end{align*}
where the weight function $W$ can be chosen in the same way as for the simple hypothesis.
%

\begin{theo}\label{theo4}
Under assumptions (M), (e) and (X2) under the null hypothesis  of centred normally distributed errors  we obtain the following expansions and asymptotic distributions, where we use the notation $\tilde \eps_j=\eps_j/\vartheta\sim N(0,1)$.
\begin{itemize}
\item[(i)] In case (ab.1') 
\begin{align*}
T_{n,C_2}
=&\ n\int\Big(\frac 1n\sum_{j=1}^n \sin\big(t\tilde\e_j\big)-E[\cos(t\tilde\e_1)]t\tilde\e_j -E[\cos(t\tilde\e_1)\tilde\e_1]\frac t2 \big(\tilde\e_j^2-1\big)\\
&\qquad+\cos\big(t\tilde\e_j\big)+E[\sin(t\tilde\e_1)]t\tilde\e_j \\
&\qquad+E[\sin(t\tilde\e_1)\tilde\e_1]\frac t2 \big(\tilde\e_j^2-1\big)-\varphi_1(t)\Big)^2W(t)dt+o_{\mathbb{P}}(1)
\end{align*}
 and
$T_{n,C_2}$ converges in distribution to $T_{C_{2,1}}=\int (G_{C_{2,1}}(t))^2 W(t)dt$ for a Gaussian process $G_{C_{2,1}}$ with covariance function
\rm
$$\Cov(G_{C_{2,1}}(s),G_{C_{2,1}}(t))=\exp\big(-\frac 12(s-t)^2\big)-\big(\frac12s^2t^2+st+1\big)\exp\big(-\frac 12(s^2+t^2)\big).$$
\it
\item[(ii)] In case (ab.2)
\begin{align*}
T_{n,C_2}
=&\ n\int\Big(\frac 1n\sum_{j=1}^n \sin\big(t\tilde\e_j\big)-E[\cos(t\tilde\e_1)\tilde\e_1]\frac t2 \big(\tilde\e_j^2-1\big)\\
&\qquad\qquad+\cos\big(t\tilde\e_j\big) +E[\sin(t\tilde\e_1)\tilde\e_1]\frac t2 \big(\tilde\e_j^2-1\big)-\varphi_1(t)\Big)^2W(t)dt+o_{\mathbb{P}}(1),
\end{align*}
and
$T_{n,C_2}$ converges in distribution to $T_{C_{2,2}}=\int (G_{C_{2,2}}(t))^2 W(t)dt$ for a centred Gaussian process $G_{C_{2,2}}$ with covariance function
\rm 
$$\Cov(G_{C_{2,2}}(s),G_{C_{2,2}}(t))=\exp\big(-\frac 12(s-t)^2\big)-\big(\frac12s^2t^2+1\big)\exp\big(-\frac 12(s^2+t^2)\big).$$
\it 
\end{itemize}

\end{theo}

\noindent {\bf Proof of Theorem \ref{theo4}.}

Expanding $\sin(t\hat\e_j/\hat\vartheta)$ we get 
\begin{align*}
\frac 1n\sum_{j=1}^n \sin\big(t\frac{\hat\e_j}{\hat\vartheta}\big) =&\ \frac 1n\sum_{j=1}^n \sin\big(t\frac{\e_j}{\vartheta}\big)-\cos\big(t\frac{\e_j}{\vartheta}\big)t(\frac{\hat\alpha}{\hat\vartheta}-\frac\alpha{\vartheta}+\langle X_j,\frac{\hat\beta}{\hat\vartheta}-\frac\beta{\vartheta}\rangle)+t^2\cdot o_{\mathbb{P}}(n^{-1/2})\\
=&\ \frac 1n\sum_{j=1}^n \sin\big(t\frac{\e_j}{\vartheta}\big)-\cos\big(t\frac{\e_j}{\vartheta}\big)\frac{t}{\vartheta}(\hat\alpha-\alpha+\langle X_j,\hat\beta-\beta\rangle)\\
&\hspace{4cm}-\cos\big(t\frac{\e_j}{\vartheta}\big)t\frac{\e_j}{\vartheta} \frac{\hat\vartheta-\vartheta}{\vartheta}+t^2\cdot o_{\mathbb{P}}(n^{-1/2}),
\end{align*}
where for the last equation we used that $(\vartheta-\hat\vartheta)=O_{\mathbb{P}}(n^{-1/2})$ as pointed out in Section \ref{gof-normal} and thus $\frac 1n\sum_{j=1}^n (\hat\e_j-\e_j)(\vartheta-\hat\vartheta)=o_{\mathbb{P}}(n^{-1/2})$ as well as $\frac 1n\sum_{j=1}^n\e_j(\vartheta-\hat\vartheta)(1/\hat\vartheta-1/\vartheta)=o_{\mathbb{P}}(n^{-1/2})$. An analogous statement holds for $\frac 1n\sum_{j=1}^n\cos(t\hat\e_j)$.

Inserting (\ref{var-expansion}) we get
\begin{align*}
&T_{n,C_2}\\
=&\ n\int\Big(\frac 1n\sum_{j=1}^n \sin\big(t\frac{\e_j}{\vartheta}\big)-\cos\big(t\frac{\e_j}{\vartheta}\big)\frac{t}{\vartheta}(\hat\alpha-\alpha+\langle X_j,\hat\beta-\beta\rangle)\\
&\qquad -E\big[\cos\big(t\frac{\e_1}{\vartheta}\big)\frac{\e_1}{\vartheta}\big]\frac t2 \big(\frac{\e_j^2}{\vartheta^2}-1\big)+\cos\big(t\frac{\e_j}{\vartheta}\big)+\sin\big(t\frac{\e_j}{\vartheta}\big)\frac{t}{\vartheta}(\hat\alpha-\alpha+\langle X_j,\hat\beta-\beta\rangle) \\
&\qquad+E\big[\sin\big(t\frac{\e_1}{\vartheta}\big)\frac{\e_1}{\vartheta}\big]\frac t2 \big(\frac{\e_j^2}{\vartheta^2}-1\big)-\varphi_1(t)\Big)^2W(t)dt+o_{\mathbb{P}}(1),
\end{align*}
 since $n\big(\frac 1n\sum_{j=1}^n\e_j^2-\vartheta^2\big)^2=\big(\frac 1{\sqrt n}\sum_{j=1}^n(\e_j^2-E[\e_j^2])\big)^2=O_{\mathbb{P}}(1)$ by the central limit theorem and $\int t^2\big(\frac 1n\sum_{j=1}^n \cos(t\e_j)\e_j-E[\cos(t\e_1)\e_1]\big)^2W(t)dt=o_{\mathbb{P}}(1)$ by an analogue of Lemma \ref{lem-3}.
 The same statement holds with cosine replaced by sine.
 
Further, we consider the case (ab.1'); the result for case (ab.2) can be derived in the same way. 
Analogously as in the simple hypothesis case we get with $\tilde\e_j=\frac{\e_j}\vartheta$
\begin{align*}
T_{n,C_2}
=&\ n\int\Big(\frac 1n\sum_{j=1}^n \sin\big(t\tilde\e_j\big)-E[\cos(t\tilde\e_1)]t\tilde\e_j -E[\cos(t\tilde\e_1)\tilde\e_1]\frac t2 \big(\tilde\e_j^2-1\big)\\
&\qquad+\cos\big(t\tilde\e_j\big)+E[\sin(t\tilde\e_1)]t\tilde\e_j \\
&\qquad+E[\sin(t\tilde\e_1)\tilde\e_1]\frac t2 \big(\tilde\e_j^2-1\big)-\varphi_1(t)\Big)^2W(t)dt+o_{\mathbb{P}}(1)
\end{align*}
using the same arguments as in (\ref{expansion-hateps}) and (\ref{expansion-3}) in the proof of Theorem \ref{theo3}. 

To derive the asymptotic covariance function note that $\tilde\e_j\sim N(0,1)$, $\varphi_1(t)=\exp(-\frac 12 t^2)=E[\cos(t\tilde\e_1)]$, $E[\sin(t\tilde\e_1)]=0$, $E[\cos(t\tilde\e_1)\tilde\e_1]=0$, $E[\sin(t\tilde\e_1)\tilde\e_1]=t\exp(-\frac 12 t^2)$.
\hfill $\Box$

\bigskip

\noindent Critical values for $T_{C_{2,1}}$ and $T_{C_{2,2}}$ calculated via Monte Carlo simulations are tabled in Tables \ref{tab5} and  \ref{tab6},  for the reader's ease of use of the tests. We reject the null hypothesis of a normal distribution if $T_{n,C_2}>c_\alpha$.

\bigskip

\begin{table}[h]
\small
\begin{tabular}{| l | l | l | l | l | l | }
\hline
nominal level  $\alpha$ &$0.15$&$0.1$&$0.05$&$0.025$&$0.01$\\
\hline
critical value $c_\alpha$ &$ 0.604$&$ 0.727 $&$ 0.938 $&$ 1.177 $&$ 1.444 $\\
\hline
\end{tabular}
\caption{ \label{tab5} \sl Asymptotic critical values for the test for Gaussian errors based on the empirical characteristic function with $W(t)=\exp(-\frac 12t^2)$ in the functional linear model under condition (ab.1').}
\end{table}
\begin{table}[h]
\small
\begin{tabular}{| l | l | l | l | l | l | }
\hline
nominal level  $\alpha$ &$0.15$&$0.1$&$0.05$&$0.025$&$0.01$\\
\hline
critical value $c_\alpha$ &$ 1.441$&$ 1.833 $&$ 2.491 $&$ 3.171 $&$ 3.881 $\\
\hline
\end{tabular}
\caption{ \label{tab6} \sl Asymptotic critical values for the test for Gaussian errors based on the empirical characteristic function with $W(t)=\exp(-\frac 12t^2)$ in the functional linear model under condition (ab.2).}
\end{table}

\section{Finite sample properties}\label{sec:simus}


We consider the Hilbert space $\mathcal{H}=L^2([0,1])$ with inner product $\langle g,h\rangle=\int_0^1 g(t)h(t)\,dt$ and norm $\|g\|=(\int_0^1 g^2(t)\,dt)^{1/2}$.
Validity of the assumption (b) under regularity conditions was shown in \cite{NeumeyerSelk2025} in Lemma 3.2 and Example 3.3. 
 We assume $\beta\in \mathcal{W}_2^m([0,1])$ for some $m>2$ and the Sobolev-space
\begin{eqnarray*}
\mathcal{W}_2^m([0,1])& &= \big\{b:[0,1]\to\mathbb{R}\mid 
b^{(j)}\mbox{ is absolutely continuous for }j=0,\dots,m-1,\\
&&\qquad\qquad\qquad\qquad\mbox{ and }\|b^{(m)}\|<\infty
\big\},
\end{eqnarray*}
where $b^{(0)}=b$ and $b^{(j)}$ denotes the $j$-th derivative of $b$, $j\geq 1$.
We consider the regularized estimators in  \cite{YuanCai2010}, i.e.
\begin{equation}\label{estimYuanCai}
\big(\hat\alpha,\hat\beta \big)=\arg\min_{a\in\mathbb{R},b\in \mathcal{W}_2^m([0,1])} \left\{\frac1n\sum_{i=1}^n 
\Big(Y_{i}- \big(a+\langle X_i,b\rangle \big)\Big)^2+\lambda_n\big\|b^{(m)}\big\|^2\right\}
\end{equation}
for a suitable positive sequence $\lambda_n$ converging to zero, which in particular gives $\hat\alpha=\overline{Y}_n-\langle \overline{X}_n,\hat\beta\rangle$, and (ab.1) is fulfilled.
The rates of convergence $|\hat\alpha-\alpha|=o_{\mathbb{P}}(n^{-1/4})$, $\|\hat\beta-\beta\|=o_{\mathbb{P}}(n^{-1/4})$ as in conditions (ab.2), (ab.3), (ab.1') can also be obtained for those estimators under regularity conditions. 

\subsection{Testing for Gaussian errors based on empirical cdf}\label{sec:sim-cdf-ab1}

For $i=1,\ldots,n$ the functional observations $X_{i}(t)$, $t \in [0,1]$, are generated according to
\begin{equation}\label{sim-model}X_{i}(t)=\frac 12\sum_{l=1}^5\Big(B_{i,l}\sin\left( t(5-B_{i,l})2\pi\right)-M_{i,l}-E[B_{i,l}\sin\left( (5-B_{i,l})2\pi\right)-M_{i,l}]\Big),\end{equation}
where $B_{i,l}\sim\mathcal{U}[0,5]$ and $M_{i,l}\sim\mathcal{U}[0,2\pi]$ for $l=1,\ldots,5$, $i=1,\ldots,n$. Here  $\mathcal{U}$ stands for the (continuous) uniform distribution. The  functional linear model is built as
$$Y_i=\int X_{i}(t)\gamma_{3,\frac 13}(t)dt+\varepsilon_i,$$
where the coefficient function $\gamma_{a,b}(t)=b^a/\Gamma(a)t^{a-1}e^{-bt}I\{t>0\}$ is the density of the Gamma distribution. Furthermore, we assume that each $X_{i}$ is observed on a dense, equidistant grid of 300 evaluation points.

The parameter estimators are the regularized estimators proposed in \cite{YuanCai2010} (see also \eqref{estimYuanCai} in the paper at hand) with $m=3$ and a data-driven tuning parameter $\lambda_n$ chosen by generalized cross-validation as described in the aforementioned paper. Because case (ab.1') applies we use the critical values tabled in Table \ref{tab1}.

To test for Gaussian errors (composite hypothesis) we consider the skew-normal error distribution with location parameter
$$-\sqrt{\frac{2\pi\left(\left(5\delta\right)^2+\left(5\delta\right)^4\right)}{\pi^2+\left(2\pi^2-2\pi\right)\cdot\left(5\delta\right)^2+\left(\pi^2-2\pi\right)\cdot\left(5\delta\right)^4}},$$
 scale parameter $\big(\pi\left(1+\left(5\delta\right)^2\right)/(\pi+(\pi-2)\cdot\left(5\delta\right)^2)\big)^{1/2}$ and shape parameter $5\delta$ for different values of $\delta$. For $\delta=0$ this is the standard normal distribution. The rejection probabilities for level $5\%$ and 500 replications are displayed in Table \ref{tab:sn}. It can be seen that the level is approximated well and the power increases for increasing parameter $\delta$ as well as for increasing sample size $n$. 
 
 As another example, we consider Student-t distributed errors with different degrees of freedom. 
The results tabled in Table \ref{tab:t} show a good power that increases for increasing sample size $n$ while it decreases for increasing parameter $\delta$, because the Student-t distribution converges to the standard normal distribution for increasing degree of freedom.
 
\begin{table}[h!]
\small
\begin{tabular}{| r || c | c | c | c | c | c | c | c |}
\hline
$\delta$ &$0$&$0.1$&$0.2$&$0.3$&$0.4$&$0.6$&$0.8$&$1$\\
\hline\hline
 $n=100$&$4.8$&$5.4$&$6.4$&$12.8$&$25.6$&$51$&74.6&81.2 \\
\hline
 $n=200$&$4.6$&$5.6$&$7.2$&$20.2$&$48$&$87$&97.4&99.4 \\
\hline
\end{tabular}
\caption{\sl Rejection probabilities for the test based on cdf in percent obtained with skew-normally distributed errors.}\label{tab:sn}
\small
\begin{tabular}{| r || c | c | c | c | c |}
\hline
$\delta$&$ 3 $&$4$&$5$&$6$&$7$\\
\hline\hline
 $n=100$&$81$&$54$&35.6&26.6&24.4 \\
 \hline
 $n=200$&96.6&85.6&64.8&49.4&39.6 \\
\hline
\end{tabular}
\caption{\sl Rejection probabilities for the test based on cdf in percent obtained with $\mathcal{S}t(\delta)$ distributed errors.}\label{tab:t}
\end{table}

\subsection{Model misspecification}

The results in Theorem \ref{theo2} show that the asymptotic distribution of the test statistic is different under condition (ab.2) than under (ab.1'). To investigate the finite sample performance under (ab.2) we generate data according to
\begin{equation*}X_{i}(t)=\frac 12\sum_{l=1}^5\Big(B_{i,l}\sin\left( t(5-B_{i,l})2\pi\right)-M_{i,l}-E[B_{i,l}\sin\left(t (5-B_{i,l})2\pi\right)-M_{i,l}]\Big),\end{equation*}
where $E[X_i(t)]=0$ $\forall t$, and adapt the estimators to the case $\alpha=\hat\alpha=0$. For this setting, the critical values from Table \ref{tab2} are used and the rejection probabilities under the null hypothesis and under the alternative of  skew-normal distributed errors for level $5\%$ and 500 replications are displayed in the left half of Table \ref{tab:miss}. It can be seen that the test is more conservative under condition  (ab.2) than under (ab.1'), but still the power increases with increasing parameter $\delta$ as well as with increasing sample size $n$. 

To investigate what happens if the model is misspecified, we again generate data according to model \eqref{sim-model}, where $E[X]\not\equiv 0$ and thus (ab.2) is not fulfilled. Nevertheless we use the critical values from Table \ref{tab2}, adapt the estimators to the case $\alpha=\hat\alpha=0$, and apply the procedure on
$$Y_i-\overline{Y}_n\approx \langle X_i-\overline{X}_n,\beta\rangle+\varepsilon_i,$$
which is a typical approach but not  recommended, as pointed out in Remark \ref{rem1}. The rejection probabilities under the null hypothesis and under the alternative of  skew-normal distributed errors for level $5\%$ and 500 replications are displayed in the right half of Table \ref{tab:miss}. It can be seen that the values are much too small, which corresponds to the findings in Remark \ref{rem1}, since for this setting the critical values from Table \ref{tab1} should have been used, which are smaller than those in Table \ref{tab2}.

\begin{table}[h!]
\small
\begin{tabular}{| r || c | c | c | c | c || c | c | c | c | c | }
\hline
&\multicolumn{4}{c}{$E[X]\equiv 0$} & & \multicolumn{4}{c}{$E[X]\not\equiv 0$} & \\
\hline
$\delta$&0&0.2&0.5&0.8& 1&0&0.2&0.5&0.8&1\\
\hline\hline
 $n=100$&$2.4$&$2.6$&6&12.8&18.4&0&0&0.2&2& 4 \\
 \hline
 $n=200$&2.6&3&14.8&30.4&41.4&0&0&2.4&20&37.6 \\
\hline
\end{tabular}
\caption{\sl Rejection probabilities for the test based on cdf in percent obtained with skew-normally distributed errors for a model that fulfills (ab.2) (left) and for a model that does not fulfill (ab.2) (right).}\label{tab:miss}
\end{table}

\subsection{Testing for Gaussian errors based on ecf}
As in Section \ref{sec:sim-cdf-ab1}, we again generate data according to model \eqref{sim-model}. To construct the test statistic, we use the empirical characteristic function introduced in Section \ref{sec:ecf}.
Apart from the test statistic, we consider exactly the same setting as described in Section \ref{sec:sim-cdf-ab1}.
Thus, condition (ab.1') is fulfilled and we use the critical values tabled in Table \ref{tab5}.

Results for skew-normal and Student-t distributed errors for level 5\%, 500 replications and weight function $W(t)=\exp(-0.5t^2)$ are displayed in Table \ref{tab:snC} and Table \ref{tab:tC}. It can be seen that the level is approximated well and the power is even higher than for the test based on the empirical cdf. The same effect is observed in \cite{HuskovaMeintanis2007}.

\begin{table}[h!]
\small
\begin{tabular}{| r || c | c | c | c | c | c | c | c |}
\hline
$\delta$ &$0$&$0.1$&$0.2$&$0.3$&$0.4$&$0.6$&$0.8$&$1$\\
\hline\hline
 $n=100$&$4.4$&$5.6$&$9.2$&$15.6$&$34.8$&$65.4$&$81.8$&$89.6$ \\
\hline
 $n=200$&$4.4$&$5.6$&$9.2$&$30.8$&$60.6$&$93.2$&$99.4$&$100$ \\
\hline
\end{tabular}
\caption{\sl Rejection probabilities for the test based on ecf in percent obtained with skew-normally distributed errors.}\label{tab:snC}
\small
\begin{tabular}{| r || c | c | c | c | c |}
\hline
$\delta$&$ 3 $&$4$&$5$&$6$&$7$\\
\hline\hline
 $n=100$&$85.2$&$63.4$&44&34.6&28 \\
 \hline
 $n=200$&97.6&86&71.6&55.8&48 \\
\hline
\end{tabular}
\caption{\sl Rejection probabilities for the test based on ecf in percent obtained with $\mathcal{S}t(\delta)$ distributed errors.}\label{tab:tC}
\end{table}

\section{Conclusion and future work}\label{concluding remarks}

In this paper we considered the error distribution in functional linear models with scalar response and functional covariate.
We proposed goodness-of-fit tests for a specific distribution of the errors (simple hypothesis) as well as for the error distribution to belong to a parametric function class (composite hypothesis). We considered test statistics based on the empirical distribution function of residuals and based on the empirical characteristic function of residuals, and derived asymptotic critical values for all proposed tests. 

Tests for symmetry of the error distribution can be based on the residual empirical distribution function with the test statistic $\sqrt{n}\sup_{z}|\hat F_n(z)-\hat F_n(-z)|$, or based on the residual empirical characteristic function $\hat\varphi_n$, testing whether it is real-valued. In the context of regression models with finite-dimensional covariates those tests have been considered by 
\cite{NeumeyerEtal2005} and 
\cite{HuskovaMeintanis2012}, and with the new results they can be applied analogously in the functional linear model under condition (ab.1'). 

Note that for the expansions derived in the paper at hand the independence of covariates $X$ and errors $\eps$ is essential. In the heteroscedastic case the expansions do not hold, and deriving asymptotic distributions will be much more complicated. To explain this consider the remainder term $R_n(z)$ in the proof of Theorem \ref{theo1}. With the notation $F_{\eps|X}(z|x)=P(\eps\leq z\mid X=x)$, condition (ab.1), and a Taylor expansion, the remainder term has the following expansion, 
\begin{eqnarray*}
R_n(z) &=& E_X[F_{\eps|X}(z+\hat\alpha-\alpha+\langle X,\hat\beta-\beta\rangle| X)-F_{\eps|X}(z|X)]\\
&\approx& \overline{\eps}_nE_X[f_{\eps|X}(z|X)]+\langle E_X[f_{\eps|X}(z|X)(X-\overline{X}_n)],\hat\beta-\beta\rangle.
\end{eqnarray*}
The second term is not $o_{\mathbb{P}}(n^{-1/2})$, and will typically even have a slower rate of convergence than $n^{-1/2}$, and will dominate the asymptotic distribution. 
Also for the empirical characteristic function, for example Lemma \ref{lem-2b} does not work in the heteroscedastic case because $E[\cos(t\eps_j)(X_j-E[X])]\neq 0$. Thus in the expansion terms based on $\hat\beta-\beta$ will dominate. 

While we consider independent observations in this paper, an extension to dependent data is interesting as well.
In \cite{Zhong2025} some tests for dependent functional data with measurement error are proposed. Their model can be interpreted as a nonparametric function-on-scalar regression. 
In future work we are planning to extend our results for the scalar-on-function regression model to the case of dependent functional observations.

\begin{appendix}

\section{Auxiliary results}

\begin{lemma}\label{lem-1} Under the assumptions of Theorem \ref{theo3} (i) and (ii) it holds that $$\frac{1}{\sqrt{n}}\sum_{j=1}^n(\hat\eps_j-\eps_j)^2=o_{\mathbb{P}}(1).$$
\end{lemma}

\noindent {\bf Proof of Lemma \ref{lem-1}.}

\noindent (i) Applying the expansion (\ref{expansion-hateps}) and Cauchy Schwarz inequality we obtain
\begin{eqnarray*}
\frac{1}{\sqrt{n}}\sum_{j=1}^n(\hat\eps_j-\eps_j)^2
&\leq& 2\sqrt{n}(\overline{\eps}_n)^2+\frac{2}{n}\sum_{j=1}^n \|X_j-E[X]\|^2\sqrt{n}\|\hat\beta-\beta\|^2\\
&&{}+2\|\overline{X}_n-E[X]\|^2\sqrt{n}\|\hat\beta-\beta\|^2\\
&=& o_{\mathbb{P}}(1)
\end{eqnarray*}
by the law of large numbers and condition (ab.1'). 

\noindent (ii) The term in the right hand side of (\ref{expansion-hateps-2}) has already been considered in case (i) and thus the assertion follows. 
\hfill $\Box$

\begin{lemma}\label{lem-2}  Under the assumptions of Theorem \ref{theo3} (i) and (ii) it holds that $$n\int \left(\frac tn\sum_{j=1}^n\cos(t\e_j)\langle \overline X_n-E[X],\hat\beta-\beta\rangle\right)^2W(t)dt=o_{\mathbb{P}}(1).$$
\end{lemma}

\noindent {\bf Proof of Lemma \ref{lem-2}.} By bounding the cosine functions and applying Cauchy Schwarz inequality we obtain the upper bound
\begin{eqnarray*}
&&n\int \left(\frac tn\sum_{j=1}^n\cos(t\e_j)\langle \overline X_n-E[X],\hat\beta-\beta\rangle\right)^2W(t)dt\\
&\leq& \int t^2 W(t)\,dt \|\sqrt{n}(\overline{X}_n-E[X])\|^2\|\hat\beta-\beta\|^2\\
&=& o_{\mathbb{P}}(1)
\end{eqnarray*}
by the central limit theorem in Hilbert space $\mathcal{H}$ and consistency of $\hat\beta$. 
\hfill $\Box$

\begin{lemma}\label{lem-2b}  Under the assumptions of Theorem \ref{theo3} (i) it holds that $$n\int \left(\frac tn\sum_{j=1}^n\cos(t\e_j)\langle X_j-E[X],\hat\beta-\beta\rangle\right)^2W(t)\,dt=o_{\mathbb{P}}(1).$$
\end{lemma}

\noindent {\bf Proof of Lemma \ref{lem-2b}.}
In this proof we use the tensor Hilbert space $\mathcal{H}\otimes \mathcal{H}$, which is a completion of the algebraic tensor product with definition of the inner product
$$\left\langle h_1\otimes h_2,b_1\otimes b_2\right\rangle_{\mathcal{H}\otimes \mathcal{H}} =
\langle h_1,b_1\rangle\cdot
\langle  h_2, b_2\rangle $$
for $h_1,h_2,b_1,b_2\in\mathcal{H}$. 
Here, as before, $\langle\cdot,\cdot\rangle$ and $\|\cdot\|$ are the notations of inner product and norm in the separable Hilbert space $\mathcal{H}$. We further use the notation $\|\cdot\|_{\mathcal{H}\otimes \mathcal{H}} $ for the norm on ${\mathcal{H}\otimes \mathcal{H}} $, which is also a separable Hilbert space. 

We use the notation 
$$g(\eps_i,\eps_j)=\int t^2 \cos(t\eps_i)\cos(t\eps_j)W(t)\,dt,$$
and obtain the expansion
\begin{eqnarray*}
&&n\int \left(\frac tn\sum_{j=1}^n\cos(t\e_j)\langle X_j-E[X],\hat\beta-\beta\rangle\right)^2W(t)\,dt\\
&=& \frac{1}{n} \sum_{i=1}^n \sum_{j=1}^n g(\eps_i,\eps_j)\left\langle (X_i-E[X])\otimes (X_i-E[X]),(\hat\beta-\beta)\otimes (\hat\beta-\beta)\right\rangle_{\mathcal{H}\otimes \mathcal{H}}\\
&=&R_{n,1}+R_{n,2}+R_{n,3},
\end{eqnarray*}
where
\begin{eqnarray*}
R_{n,1} &=& \Big\langle \frac{1}{n} \sum_{i=1}^n \sum_{j=1\atop i\neq j}^n ( g(\eps_i,\eps_j)-E[g(\eps_1,\eps_2)]) (X_i-E[X])\otimes (X_j-E[X]),\\
&&\qquad\qquad\qquad(\hat\beta-\beta)\otimes (\hat\beta-\beta)\Big\rangle_{\mathcal{H}\otimes \mathcal{H}}\\
R_{n,2} &=& E[g(\eps_1,\eps_2)] \frac{1}{n} \sum_{i=1}^n \sum_{j=1}^n \left\langle (X_i-E[X])\otimes (X_j-E[X]),(\hat\beta-\beta)\otimes (\hat\beta-\beta)\right\rangle_{\mathcal{H}\otimes \mathcal{H}}\\
&=& E[g(\eps_1,\eps_2)] \left(\left\langle\frac{1}{\sqrt{n}} \sum_{i=1}^n  (X_i-E[X]),\hat\beta-\beta \right\rangle\right)^2\\
R_{n,3} &=& \frac{1}{n} \sum_{i=1}^n (g(\eps_i,\eps_i)-E[g(\eps_1,\eps_2)])\Big\langle (X_i-E[X])\otimes (X_i-E[X]),\\
&&\qquad\qquad\qquad(\hat\beta-\beta)\otimes (\hat\beta-\beta)\Big\rangle_{\mathcal{H}\otimes \mathcal{H}}\\
&=&\frac{1}{n} \sum_{i=1}^n (g(\eps_i,\eps_i)-E[g(\eps_1,\eps_2)])\left(\langle X_i-E[X],\hat\beta-\beta\rangle\right)^2.
\end{eqnarray*}
Then by definition of $g$ and Cauchy Schwarz inequality 
$$|R_{n,2}|\leq \int t^2 W(t)\,dt \left\|\frac{1}{\sqrt{n}} \sum_{i=1}^n  (X_i-E[X])\right\|^2 \|\hat\beta-\beta\|^2=o_{\mathbb{P}}(1),$$
and 
$$|R_{n,3}|\leq 2\int t^2 W(t)\,dt \frac{1}{n} \sum_{i=1}^n \left\| X_i-E[X]\right\|^2 \|\hat\beta-\beta\|^2=o_{\mathbb{P}}(1).$$
Further by Cauchy Schwarz inequality
\begin{eqnarray*}
|R_{n,1}| &\leq& 
\left\| 
\frac{1}{n} \sum_{i=1}^n \sum_{j=1\atop i\neq j}^n 
( g(\eps_i,\eps_j)-E[g(\eps_1,\eps_2)]) 
(X_i-E[X])\otimes (X_j-E[X])
\right\|_{\mathcal{H}\otimes \mathcal{H}}\\
&&{} 
\cdot\|(\hat\beta-\beta)\otimes (\hat\beta-\beta)
\|_{\mathcal{H}\otimes \mathcal{H}}\\
&=& \left\| \sqrt{n} U_n
\right\|_{\mathcal{H}\otimes \mathcal{H}} o_{\mathbb{P}}(1),
\end{eqnarray*}
where we have applied $\|(\hat\beta-\beta)\otimes (\hat\beta-\beta)
\|_{\mathcal{H}\otimes \mathcal{H}}=\|\hat\beta-\beta\|^2=o_{\mathbb{P}}(n^{-1/2})$ by condition (ab.1'), and $U_n$ is a $\mathcal{H}\otimes \mathcal{H}$-valued U-statistic
$$U_n=
\frac{1}{n(n-1)} \sum_{i=1}^n \sum_{j=1\atop i\neq j}^n 
k(Z_i,Z_j)$$
with kernel 
$$k(Z_i,Z_j)=( g(\eps_i,\eps_j)-E[g(\eps_1,\eps_2)]) 
(X_i-E[X])\otimes (X_j-E[X])$$
and $Z_i=(X_i,\eps_i)$, $i=1,\dots,n$. Note that $E[k(Z_1,Z_2)]=0$ and 
$$E[\|k(Z_1,Z_2)\|^2_{\mathcal{H}\otimes \mathcal{H}}]\leq 4\left(\int t^2W(t)\,dt\right)^2E\|X_1-E[X]\|^4 <\infty$$ by assumption (X2). Thus $\|\sqrt{n}U_n\|_{}=O_{\mathbb{P}}(1)$ by applying the central limit theorem for U-statistics with values in separable Hilbert spaces, see \cite{PuriSazonov1991}, and we obtain $|R_{n,1}|=o_{\mathbb{P}}(1)$.
\hfill $\Box$

\begin{lemma}\label{lem-3}   Under the assumptions of Theorem \ref{theo3} (i) and (ii) it holds that
 $$\int t^2\left(\frac 1n\sum_{j=1}^n \cos(t\e_j)-E[\cos(t\e_1)]\right)^2W(t)dt=o_{\mathbb{P}}(1).$$ 
\end{lemma}
 
\noindent {\bf Proof of Lemma \ref{lem-3}.}
Note that 
\begin{eqnarray*}
&&\int t^2\left(\frac 1n\sum_{j=1}^n \cos(t\e_j)-E[\cos(t\e_1)]\right)^2W(t)dt\\
&=& \frac{1}{n^2}\sum_{i=1}^n \sum_{j=1\atop j\neq i}^n \int t^2 (\cos(t\e_i)-E[\cos(t\e_1)])(\cos(t\e_j)-E[\cos(t\e_1)]) W(t)\,dt\\
&&{}+ \frac{1}{n^2}\sum_{j=1}^n \int t^2 (\cos(t\e_i)-E[\cos(t\e_1)])^2 W(t)\,dt\\
&=& o_{\mathbb{P}}(1),
\end{eqnarray*}
which follows from the laws of large numbers for U-statistics and for arithmetic means.\hfill $\Box$

\end{appendix}

\bibliography{Bib}

\end{document}